# Simulation of clustering algorithms in OODBs in order to evaluate their performances

**J. Darmont** **A. Attoui** **M. Gourgand**

Université Blaise Pascal – Clermont-Ferrand II
Laboratoire d'Informatique
Complexe scientifique des Cézeaux
63177 Aubière Cedex
France

E-mail : darmont@libd1.univ-bpclermont.fr

**Abstract:** *A good object clustering is critical to the performance of object-oriented databases. However, it always involves some kind of overhead for the system. The aim of this paper is to propose a modelling methodology in order to evaluate the performances of different clustering policies. This methodology has been used to compare the performances of three clustering algorithms found in the literature (Cactis, CK and ORION) that we considered representative of the current research in the field of object clustering. The actual performance evaluation was performed using simulation. Simulation experiments showed that the Cactis algorithm is better than the ORION algorithm and that the CK algorithm totally outperforms both other algorithms in terms of response time and clustering overhead.*

## 1. INTRODUCTION

In Object-Oriented Databases (OODBs), a good object clustering is critical to performance [TSAN91]. Clustering means storing related objects close together on secondary storage so that when one object is accessed from disk, all its related objects are also brought into memory. Then access to these related objects is a main memory access that is much faster than a disk access.

Several methods can be used to evaluate the performances of a Database Management System (DBMS). Benchmarks generally propose a standard database and a series of operations that run on this database. Thus, performance measurement directly depends on the reactions of the tested DBMS. Several benchmarks have been specifically designed for object-oriented databases, like the Synthetic Benchmark [KIM90], the HyperModel Benchmark [ANDE90, BERR91], the OO1 Benchmark [CATT91] or the CluB-0 Benchmark [BANC92]. However, some OODB designers or clustering algorithm authors prefer the use of simulation [CHAN89, CHEN91, HE93], because simulation allows to specifically measure performance improvements due to one or another clustering policy. [TSAN92] proposes a dual performance evaluation method, performing simulations that use the database introduced by the CluB-0 Benchmark. One last way to determine the advantages of a given clustering method is mathematical analysis as it is performed by [CHAB93]. This approach is however limited because the obtained results are qualitative rather than quantitative and sharp performance criteria cannot be extracted.

The aim of this paper is to propose a methodology in order to compare the performance of the different clustering strategies that can be implemented in OODBs. Modelling may lead either to simulation or to the application of exact analytical methods whenever possible. We applied our methodology to three object-oriented clustering algorithms that are representative of the current research in the field of OODBs: Cactis [HUDS89], CK [CHAN90] and ORION [BANE87, KIM90].

The main advantage of our approach opposed to the use of benchmarks is that it allows, by providing a common environment, to specifically compare clustering algorithms, in a way that is totally independent of any environment associated with the DBMSs that implement the clustering algorithms we intend to compare. For instance, physical storage methods and buffering strategies also influence the DBMS global performance. Furthermore, our approach also allows to *a priori* study the behavior of algorithms (like CK) that are not implemented in any DBMS. Thus we can compare their performances to those of already implemented algorithms.

This paper is organized as follows. We start by presenting the modelling methodology we used. Section 3 is dedicated to our study: we apply our modelling methodology to obtain a *knowledge model* and an *action model*. Then we present in Section 4 the three studied clustering algorithms. The simulation results are given in Section 5. They expand those already provided by [DARM95]. We end this paper with a conclusion and a brief discussion about future research directions.

## 2. MODELLING METHODOLOGY

OODBs are complex systems. Modelling their behavior may as well be a complex task. This is the reason why we propose an approach dedicated to the study of such systems. This modelling approach carry out a model according to an iterative process. [GOUR91]. This process is divided into four phases (Figure 1).

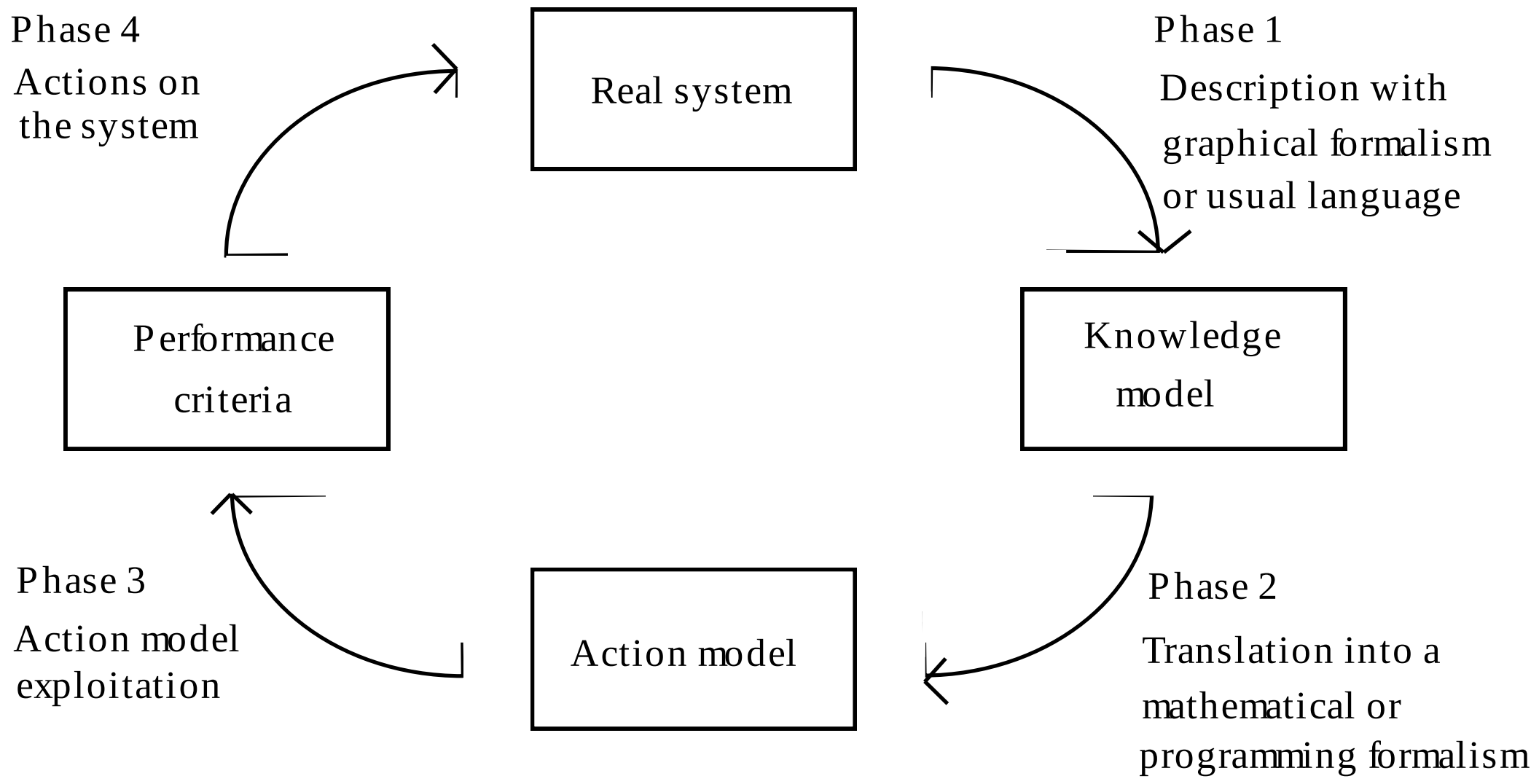

**Figure 1. Modelling iterative process**

- *Phase 1:* Analysis and formalizing of data. This system specification phase leads to the design of the *knowledge model*. Designing the knowledge model is a crucial step in the modelling process.
- *Phase 2:* Translation of the knowledge model into an *action model* using a formalism allowing its exploitation to provide performance criteria.
- *Phase 3:* Exploitation of the action model to provide performance criteria.
- *Phase 4:* Results interpretation and decisions about actions to perform on the system.

The analysis approach of a system in order to model it is performed through several steps:

- decomposition of the system to identify the different levels;
- decomposition of the system into three subsystems;
- *logical subsystem* specification;
- *physical subsystem* specification;

- *decision subsystem* specification;
- specification of the communications between the three subsystems.

*Note:* The system analysis must be iterative so that the same level of detail is achieved for all the subsystems.

## 3. STUDY

We present in this section the application of the methodology we introduced in the previous section to the domain of object-oriented databases. Though we focus on the efficiency of clustering strategies, we do not make any reference in this section to any precise clustering algorithm.

### 3.1 Domain analysis

Proceeding to a domain analysis is equivalent to designing a knowledge model of the studied system domain. Domain analysis as we performed it specifies the different entities that characterize object-oriented databases. It is shown by Figure 2 as an entity-relationship (E/R) model [CHEN76]. We could have preferred to E/R a more sophisticated semantic model , such as OMT or OOA; but while simple, E/R provides a description capability that is particularly adapted to our needs (that are limited to flat representations). Furthermore, the E/R model's relative simplicity greatly helps the dialogue between DBMSs designers or users and modelling experts. Eventually, translating an E/R model into an object-oriented model is in most cases not difficult.

The object-oriented approach presents several advantages in the field of modelling. In addition to the usual advantages of object-oriented approaches (they are natural and reliable approaches, etc.), it is particularly sensible in the field of modelling that the object-oriented approach is unifying because, by using the same approach (or even the same formalism) at all the modelling process levels, communications between the modeller and the modelled system experts is made a lot easier. Furthermore, transitions between one step of the analysis and the next step is also made easier by the use of same concepts and notations.

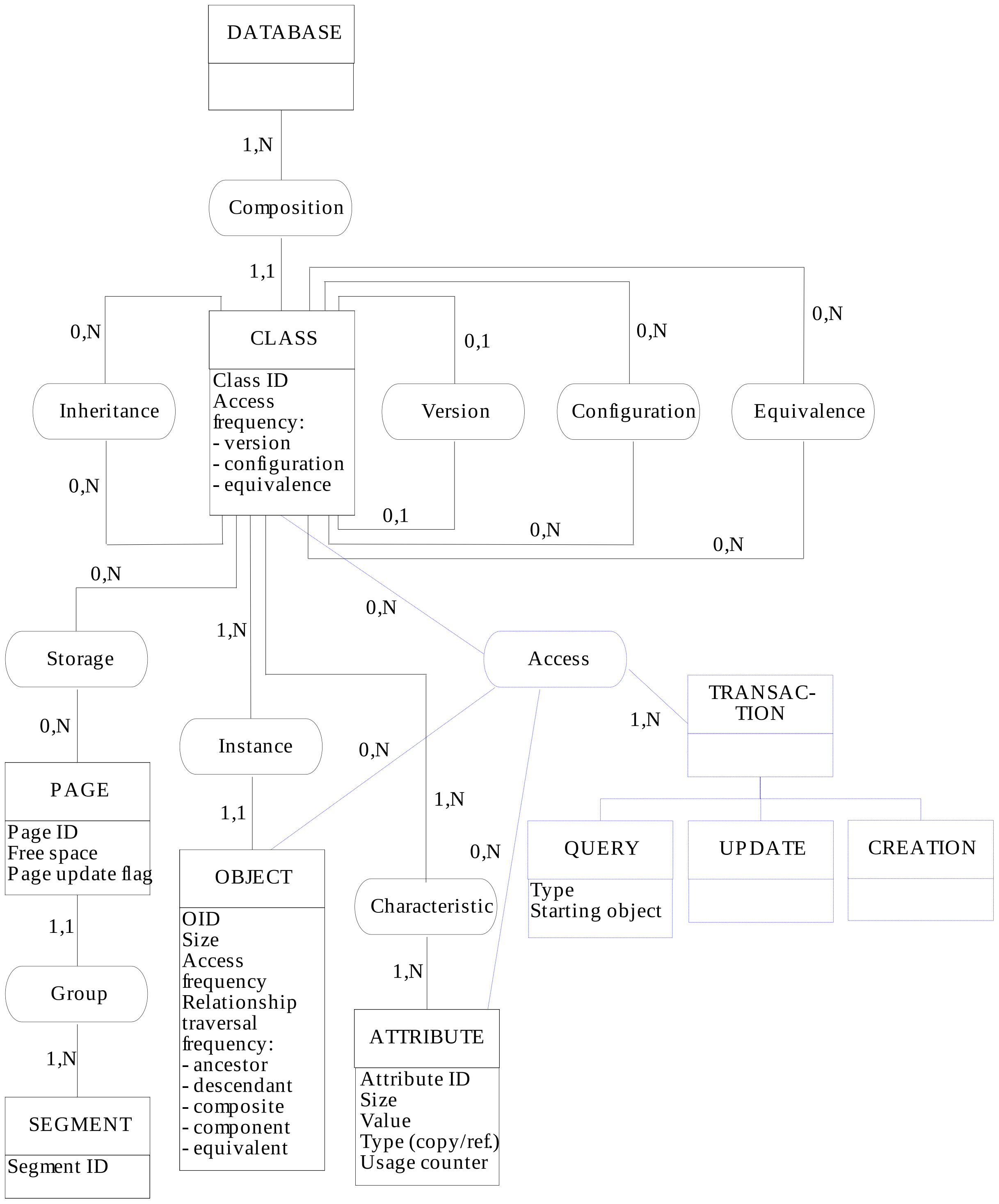

**Figure 2. Domain analysis**

## 3.2 Knowledge model

We need to describe in our model the execution of transactions on an object-oriented database. We assimilated those transactions to flows running through a system and thus designed a knowledge model using the SADT actigrams formalism (Figures 3, 4, 5, 6).

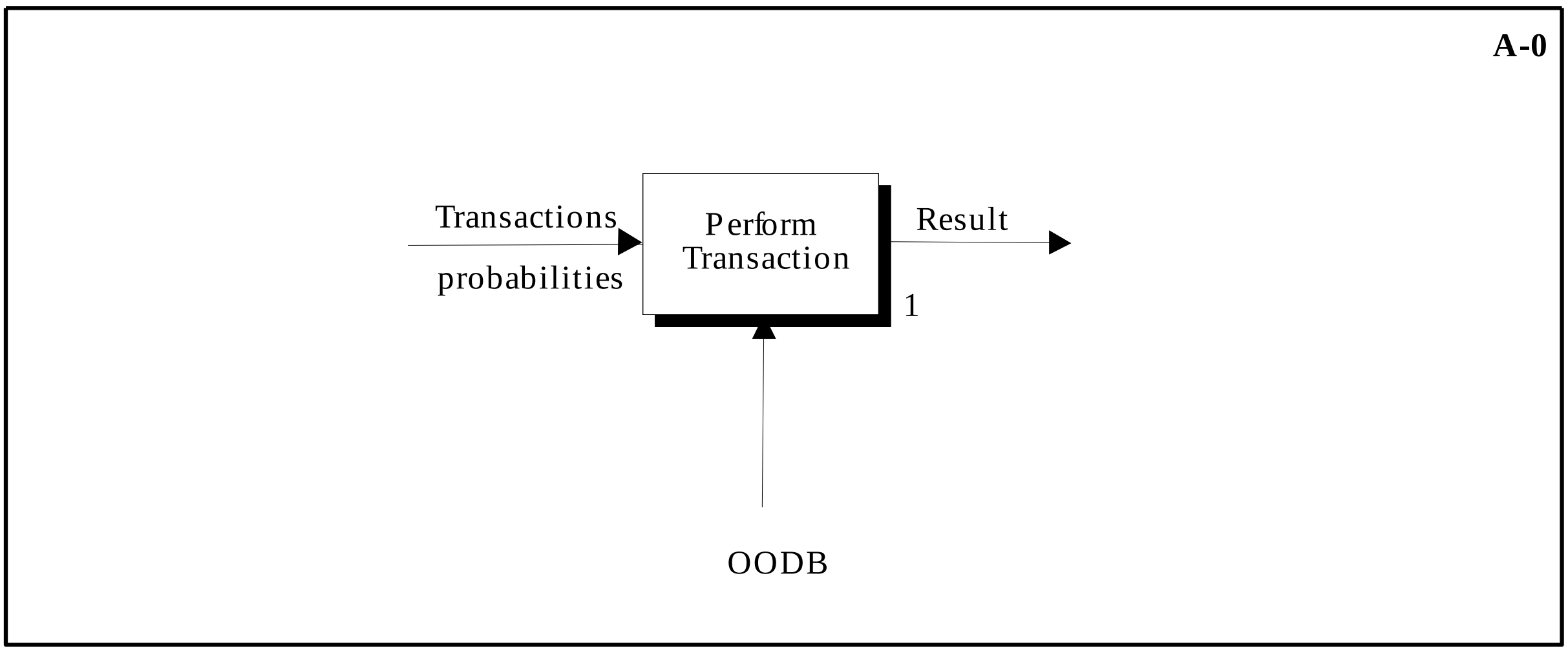

**Figure 3. Knowledge model (level A-0)**

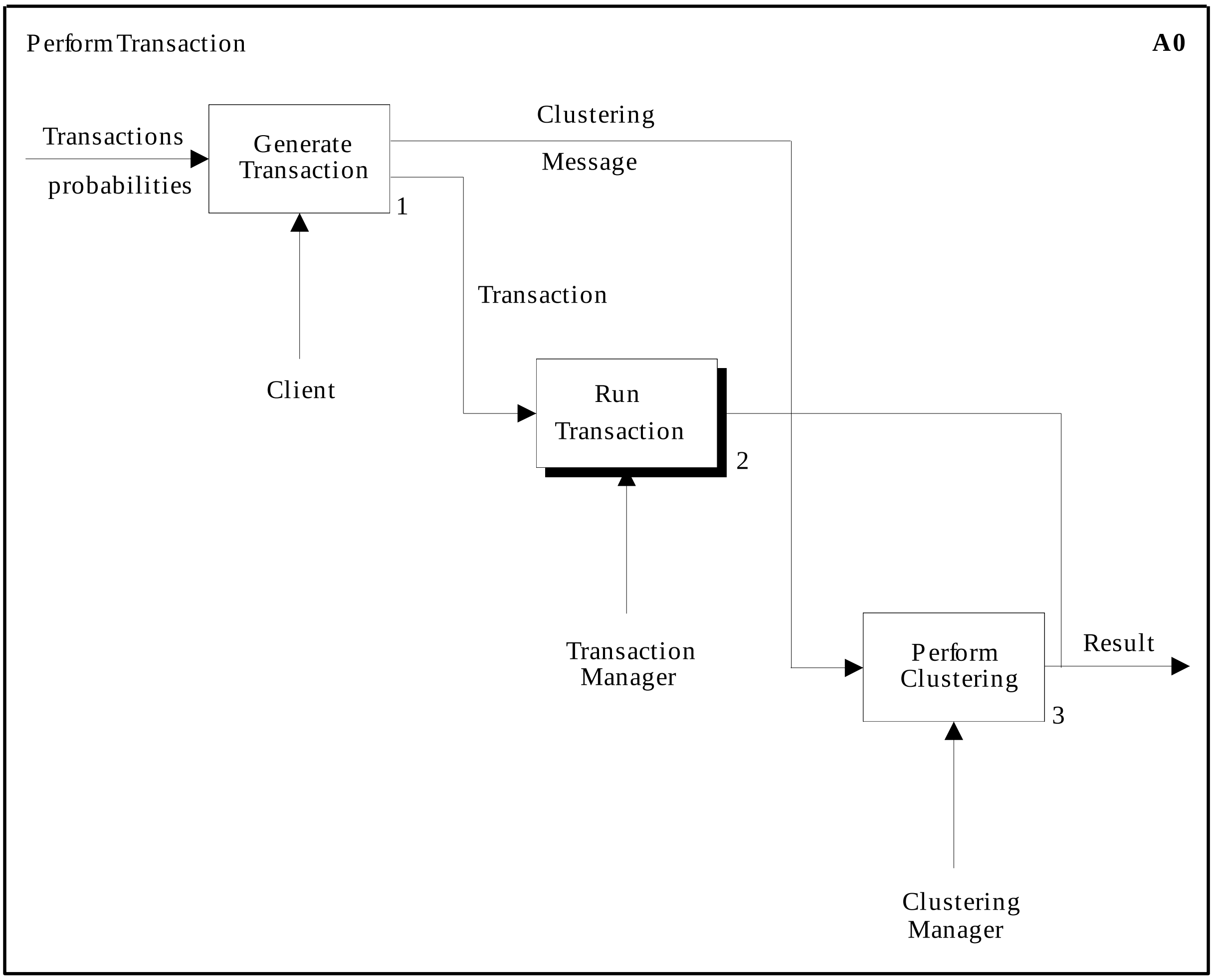

**Figure 4. Knowledge model (level A0)**

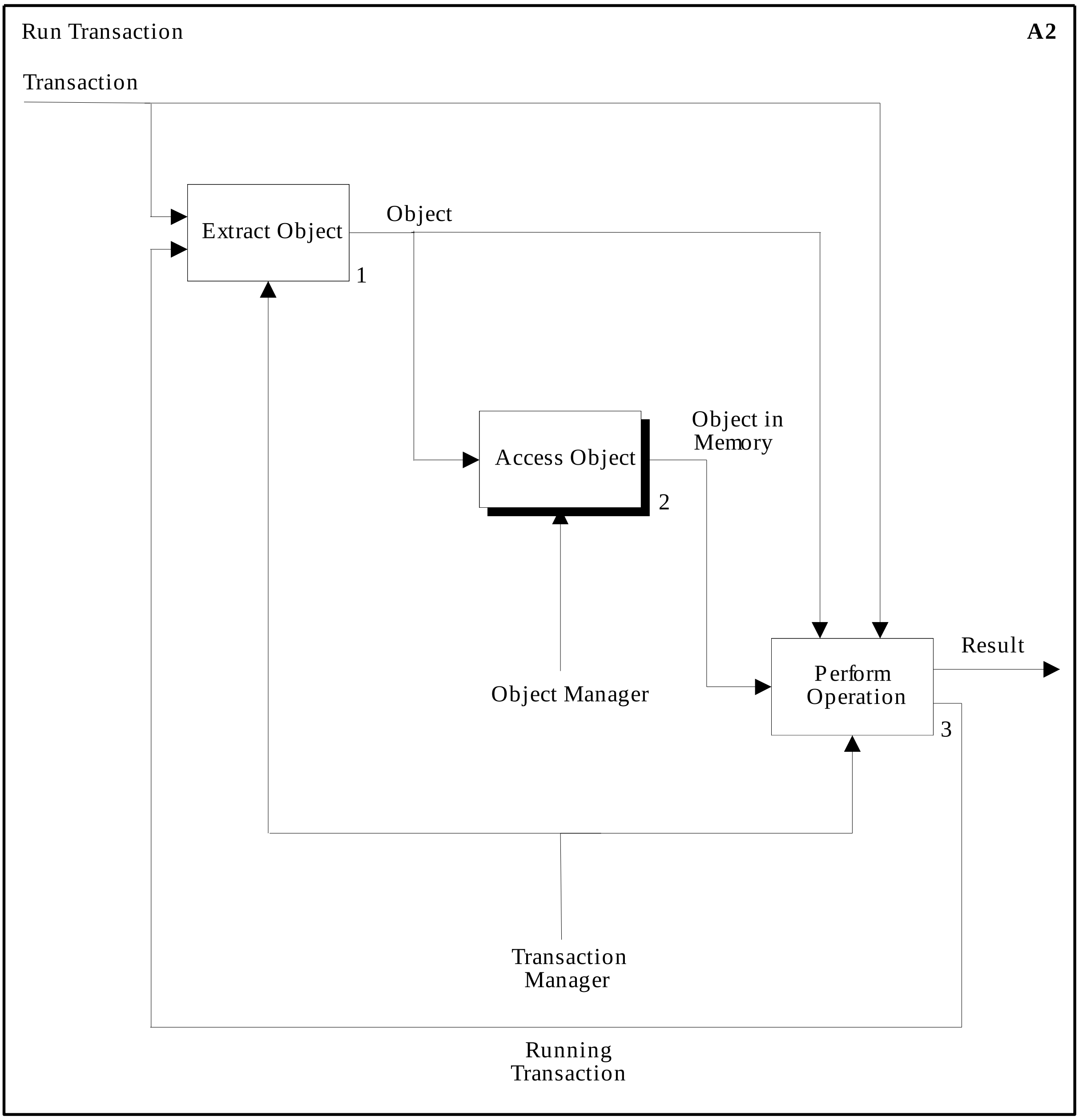

**Figure 5. Knowledge model (level A2)**

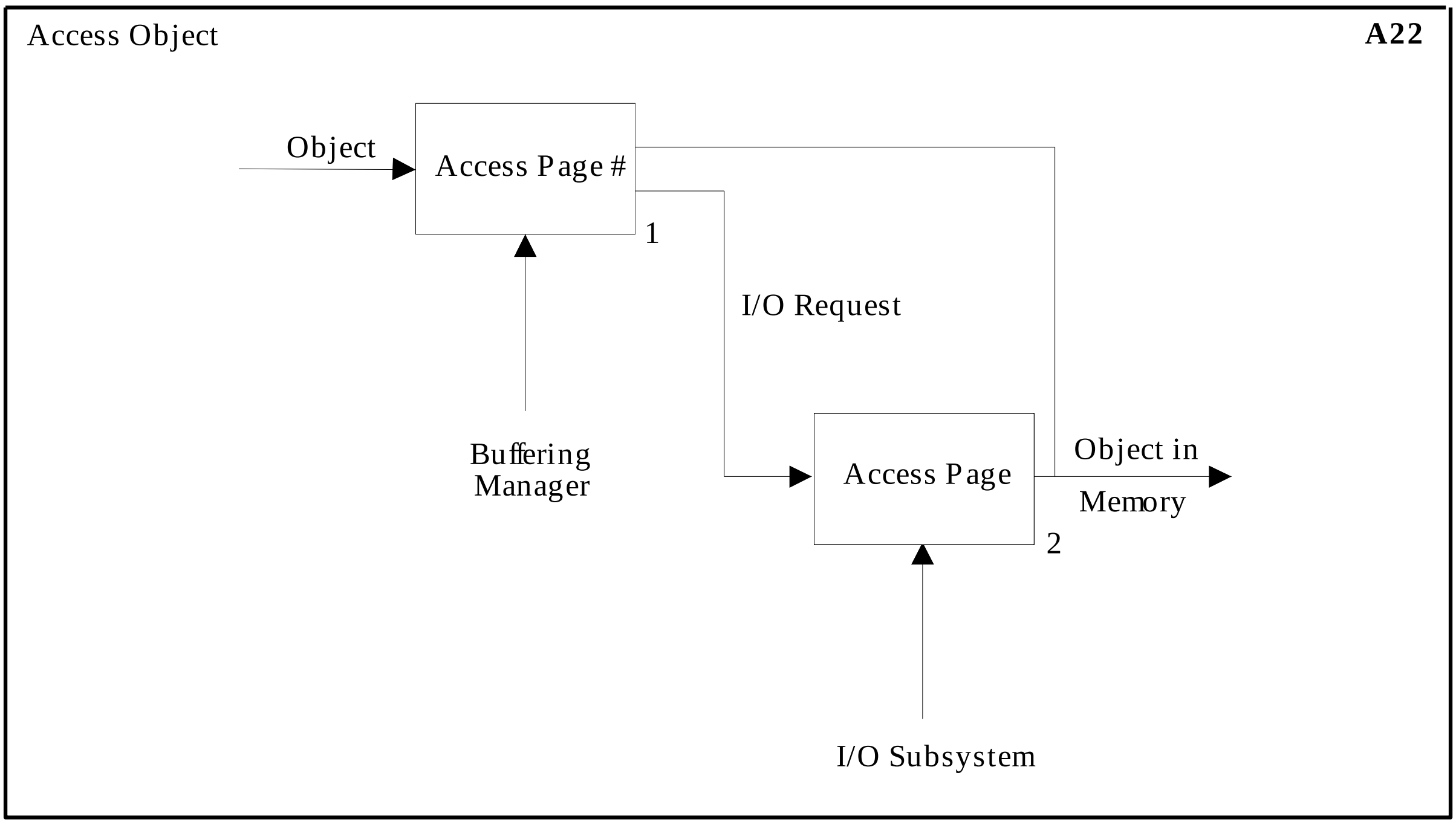

**Figure 6. Knowledge model (level A22)**

### *3.2.1 Logical subsystem*

The logical subsystem specifies what are the flows running through the system. In the case of DBMSs, these flows are *transactions* flows. The transactions are described on two levels: first, their type and then the steps of their execution. The HyperModel Benchmark [ANDE90, BERR91] provides 20 different types of transactions. From those 20, we have isolated and used 15 types of transactions (Figure 7).

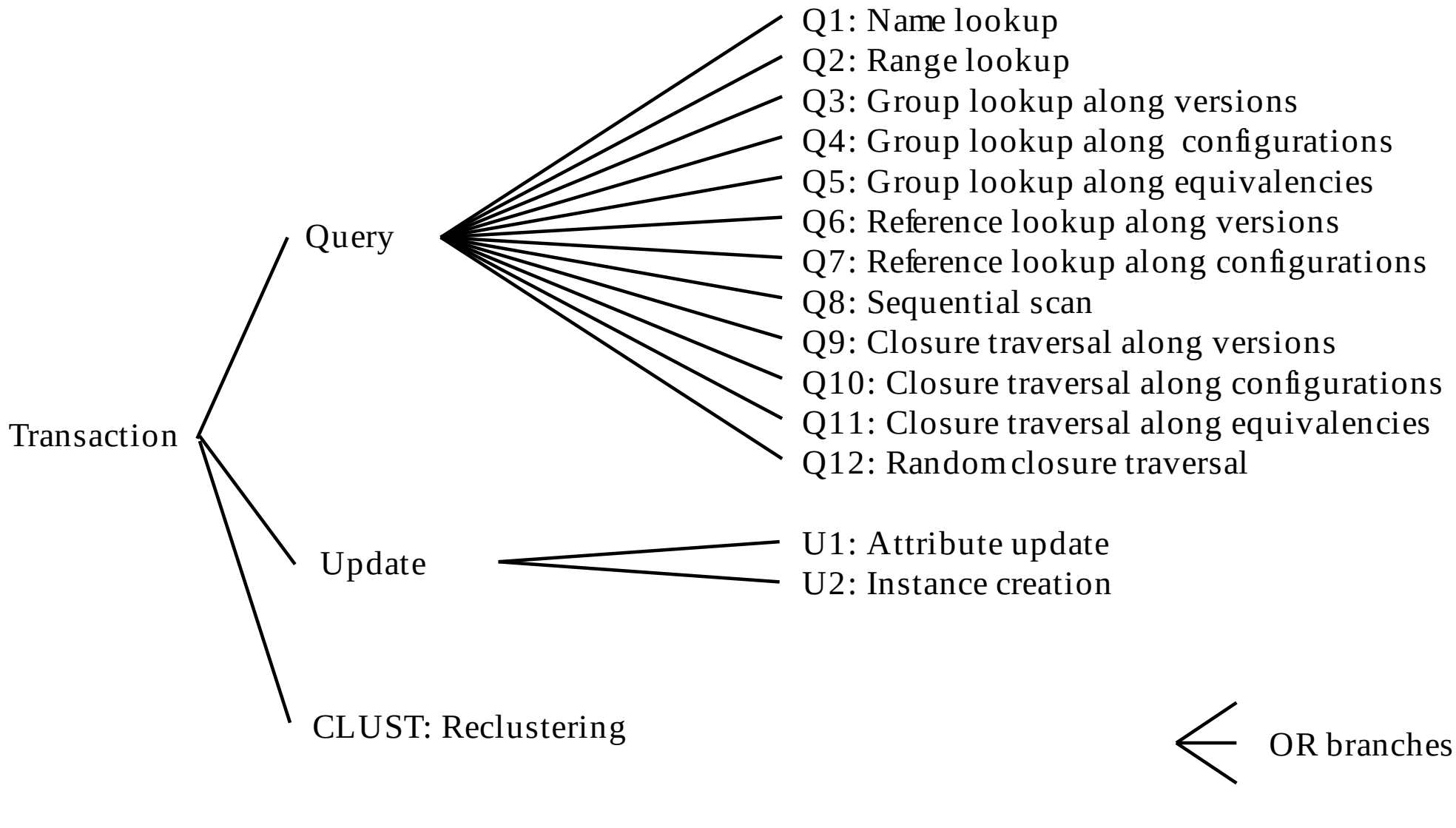

**Figure 7. Transactions' types**

- *Name Lookup*: Retrieve a randomly selected object.
- *Range Lookup*: Fetch all the instances of a given class such that a given attribute value is in a given range.
- *Group Lookup*: Given a randomly selected starting object, fetch all its descendant versions (Group lookup along versions), all its component objects (Group lookup along configuration) or all its equivalent objects (Group lookup along equivalencies).
- *Reference Lookup*: It is a "reverse" group lookup. Given a randomly selected starting object, fetch either all its ancestor versions (Reference lookup along versions) or its composite object (Reference lookup along configurations).
- *Sequential scan*: Fetch all the instances of a given class.
- *Closure Traversal*: Given a randomly selected starting object, follow one of the three structural relationships (i.e., version, configuration or equivalence) to a certain predefined depth; fetch the resulting object; the followed relationship can be either always the same (Closure traversal along versions, configurations or equivalencies) or randomly selected (Random closure traversal).

The different steps in the execution of the transactions include the following operations:

- select an object to access,
- access to the page number of the disk page containing an object,
- read or write a page on disk (i.e., perform an I/O),
- access to the attributes of an object,
- update an attribute value,
- place an object in a disk page.

### *3.2.2 Physical subsystem*

The physical resources that make up the physical subsystem are divided into two categories: *active resources* that perform some task and *passive resources* that do not directly participate into any treatment but are used by the active resources to perform their operations.

Our model active resources follow:

- AR1: User (transactions generation);
- AR2: Transactions manager (transactions execution);
- AR3: Object manager (access to objects);
- AR4: Buffering manager (application of a buffering strategy);
- AR5: I/O subsystem (disk accesses to pages);
- AR6: Clustering manager (implementation of one of the clustering algorithms that we want to evaluate).

Physical passive resources are the following:

- PR1: Main processor,
- PR2: Main memory,
- PR3: I/O processor and disk(s).

We added a fourth passive resource (PR4: Scheduler) intended to apply a scheduling policy for the transactions.

### *3.2.3 Decision subsystem*

The decision subsystem specifies what are the functioning or supervision rules in the DBMS. Each *decision rule* listed below as examples (Table 1) is associated to an SADT activity and is also a method of an object identified in the domain analysis.

| Rule code | Rule designation | Method of object |
|---|---|---|
| R1 | Generate transaction | Transaction |
| R2.1 | Extract object | Transaction |
| R2.2.1 | Access page # | Object |
| R2.2.2 | Access page | Page |
| R2.3 | Perform operation | Attribute |
| R3 | Perform clustering | Database |

**Table 1. Decision rules list**

- Rule R1 is the transaction random generation by User (AR1). These transactions are then submitted to the Transaction manager (AR2).
- Rule R2.1 is the extraction of the objects to access according to the transaction type. It is executed by the Transaction manager (AR2).
- Rule R2.2.1 is the access by the Buffering manager (AR4), via several hash tables, to the disk page number of the page containing an object to access.
- Rule R2.2.2 is the execution of an I/O performed by the I/O subsystem (AR5).
- Rule R2.3 is the execution of an operation proper to a transaction and concerning the attributes of the accessed objects by the Transaction manager (AR2).
- Rule R3 represents the execution of an object reclustering by the Clustering manager (AR6).

### 3.3 Action Model

We first translated our knowledge model in a generic action model. Tables 2 and 3 provide the simulation parameters we used for our simulation experiments. After being validated, the generic action model has been instanced for each tested clustering algorithm.

| Parameter name | Designation | Value | References |
|---|---|---|---|
| MULTI | Multiprogramming level | 10 | [GRUE91] |
| WDSIZE | Memory word size | 4 bytes | [GRUE91] |
| PGSIZE | Disk page size | 2048 bytes | [CHEN91] |
| MINTER | Mean time between two transaction generations | 4 s | [CHAN89] |
| CCT | Time necessary for concurrency control | 0.5 ms | [SRIN91] |
| ACCM | Memory word access time | 0.0001 ms | [GRUE91] |
| TEST | Testing time (in memory) | 0.0007 ms | [GRUE91] |
| SEEK | Disk seek time | 28 ms | [CHEN91] |
| LATENCY | Disk latency time | 8.33 ms | [CHEN91] |
| TRANSFER | Disk transfer time | 1.28 ms | [CHEN91] |

**Table 2. Static parameters**

| Parameter name | Designation | Default value | Range |
|---|---|---|---|
| NCL | Number of classes | 20 | 10-30 |
| NOBJ | Initial number of objects | 400 | 100-1000 |
| MNVER | Mean number of versions of a class | 3 | 1-5 |
| MNATTR | Mean number of attributes | 10 | 5-20 |
| MSATTR | Mean size for an attribute | 1 word | 1-3 words |
| BUFSIZE | Buffer size | 10 pages | 10-100 pages |
| MAXDEPTH | Maximum depth in closure traversals | 5 | 3-10 |
| PSUPER | Probability of having a superclass | 0.9 | 0-1 |
| PCOMP | Probability of being a component class | 0.5 | 0-1 |
| PEQUI | Probability of having an equivalent class | 0.1 | 0-1 |
| PQ1-PQ12 | Query probabilities | 0.065 | 0-1 |
| PU1 | Attribute update probability | Depends on tested algorithm | 0-1 |
| PU2 | Instance creation probability | 0.05 | 0-1 |
| PCLUST | Reclustering probability | Depends on tested algorithm | 0-1 |

**Table 3. Dynamic parameters**

To implement our action model (here, a simulation model), we used the QNAP2 (Queuing Network Analysis Package 2nd generation) software, version 9.0. We selected this simulation language for several reasons:

- QNAP2 is a validated simulation tool;
- QNAP2 allows the use of an object-oriented approach (since version 6.0);
- QNAP2 algorithmic language (derived from PASCAL) allows a relatively easy implementation of such complex algorithms as clustering algorithms.

Our actual simulation model (Figure 8) is built as follows.

- *User module:* After a predefined think time, the user issues the transactions to the Transaction Manager according to some frequencies of occurrence.
- *Transaction Manager module:* The Transaction Manager extracts from transactions which objects need to be accessed or updated, and performs the operations. In the case of a query (Q1-Q12) or update (U1) operation, object requests are sent to the Object Manager. In the case of instance creation (U2) or reclustering (CLUST), the Clustering Manager is invoked.
- *Object Manager module:* The Object Manager extracts the disk page the object belongs to, and then requests it to the Buffering Manager.
- *Buffering Manager module:* The Buffering Manager checks if a page is in main memory and requests it to the I/O Subsystem if it is not. It also deals with page replacement strategies. We used the following voluntarily simplistic policy: when a new page is needed, the oldest page in memory is dropped and replaced by the new one.
- *Clustering Manager module:* The Clustering Manager is activated depending on the algorithm (i.e., Cactis, CK or ORION) it implements. It deals with reorganizing the database on secondary storage to achieve better performance.
- *I/O Subsystem module:* This module deals with physical accesses to secondary storage.

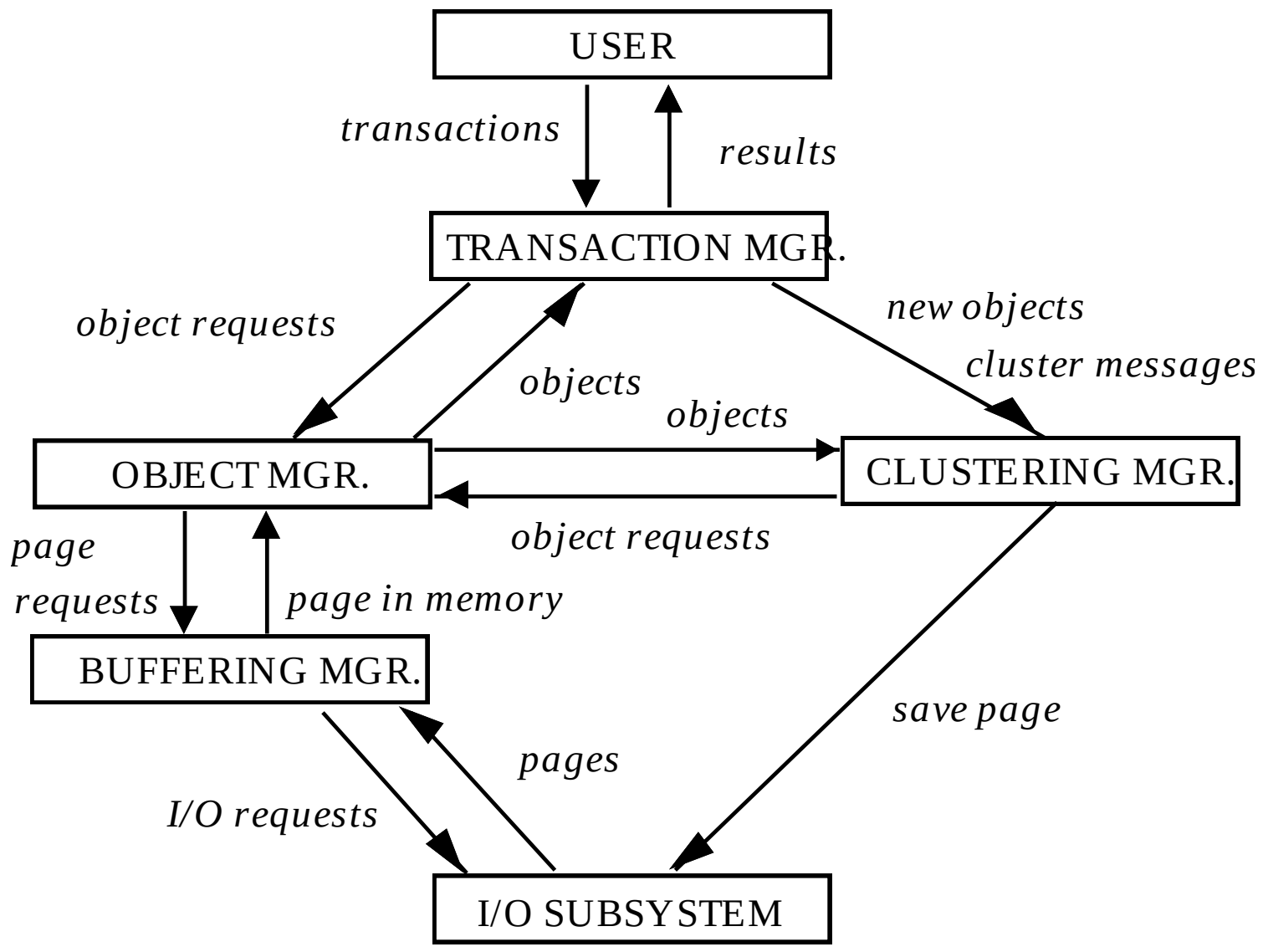

**Figure 8. QNAP2 simulation model structure**

## 4. STUDIED CLUSTERING ALGORITHMS PRESENTATION

### 4.1 Cactis [HUDS89]

Cactis is an object-oriented, multi-user DBMS developed at the University of Colorado. It is designed to support applications that require rich data modelling capabilities and the ability to specify functionally-defined data.

The Cactis clustering algorithm is designed to place objects that are frequently referenced together into the same block (i.e., page, i.e., I/O unit) on secondary storage. In order to improve the locality of data references, data is clustered on the basis of usage patterns. A count of the total number of times each object in the database is accessed is kept, as well as the number of times each relationship between objects in the process of attribute evaluation or marking out-of-date is crossed. Then, the database is periodically reorganized on the basis of this information. The database is packed into blocks using the greedy algorithm shown in Figure 9.

This clustering algorithm is also implemented in the Zeitgeist system [FORD88].

```
Repeat
    Choose the most referenced object in the database that has not yet been assigned a block.
    Place this object into a new block.
    Repeat
        Choose the relationship belonging to some object assigned to the block such that:
            (1) the relationship is connected to an unassigned object outside the block and,
            (2) the total usage count for the relationship is the highest.
        Assign the object attached to this relationship to the block.
    Until the block is full.
Until all objects are assigned blocks.
```

**Figure 9. Cactis clustering algorithm** [HUDS89]

### 4.2 ORION [BANE87, KIM90]

ORION is a series of next-generation database systems that have been prototyped at MCC (Microelectronics Computer Technology Corp.) as vehicles for research into the next-generation database architecture and into the integration of programming languages and databases. ORION has been designed for Artificial Intelligence (AI), Computer-Aided Design and Manufacturing (CAD/CAM) and Office Information System (IOS) applications.

ORION supports only a simple clustering scheme. Instances of the same class are clustered in the same physical segment (i.e., a number of blocks or pages). Each class is associated with one single segment.

Composite objects may also be clustered in multi-classes segments. User assistance is required to determine which classes should share the segment. The user can dynamically issue a Cluster message containing a “ListOfClassNames” argument specifying the classes that are to be placed in the same segment.

### 4.3 **CK** [CHAN90]

The CK algorithm (from its authors' names: Chang and Katz) is defined in the CAD/CAM context. It is not yet implemented in any OODB.

The CK algorithm is based on a particular inheritance link called *instance-to-instance* and *inter-objects access frequencies* (given by the user at data type creation time) for each kind of structural relationship (i.e., versions, configurations and equivalencies). These access frequencies and a computation of the costs of instance-to-instance inherited attributes give the page where a new object has to be placed. [BULL95]

The concept of instance-to-instance inheritance is an extension of the classical inheritance relationship (the IS-A relationship). Instance-to-instance inheritance not only transfers the existence of attributes from one object to another (like type inheritance), but moreover the values of these attributes. For example, instance-to-instance inheritance is important in computer-aided design databases, since a new version tends to resemble its immediate ancestor. It is useful if a new version can inherit its attributes values, and more importantly its constraints, from its ancestor.

The pseudo code of the CK algorithm is provided in the appendix.

## 5. SIMULATION RESULTS

### 5.1 Performance criteria

- The first performance criteria we came up with is the *mean response time*. It is a good metric for overall performance. Response time is measured for each type of transaction and takes into account the clustering overhead in the case of queries and updates.
- We also measured the *mean number of I/Os*, that we further divided into two categories. Transactions I/Os is the number of I/Os performed to complete regular transactions (i.e., queries and updates). Transactions I/Os may be an indication on how well objects are clustered. Clustering I/O overhead directly gives clustering overhead.

- Storage space is a crucial parameter when speaking of databases. Thus we measured the *mean number of disk pages* necessary to each algorithm to cluster the database.
- We last selected the *mean system throughput* as a performance criteria. However, it appeared after our simulation experiments that this criteria was not significant since the average transaction execution time is far less than the mean time between two transaction generations (4 seconds). Hence, the system throughput was always close to the optimal (0.25 transaction per second) and did not vary much.

## 5.2 Results

### *5.2.1 Effects of the database initial size*

Database size directly influence DBMSs performances, and in particular clustering algorithms performances. In this series of simulations, we varied the database initial size, i.e., the database size before simulation (before new instances are created).

Mean response time for each clustering algorithm is given by Figure 10. Two graphs are necessary because each of them use a very different scale. Figure 10 shows indeed that Cactis is better than ORION (2.5 times better). The CK algorithm performances are far greater than those of Cactis and ORION (they are 1,100 times better that those of Cactis).

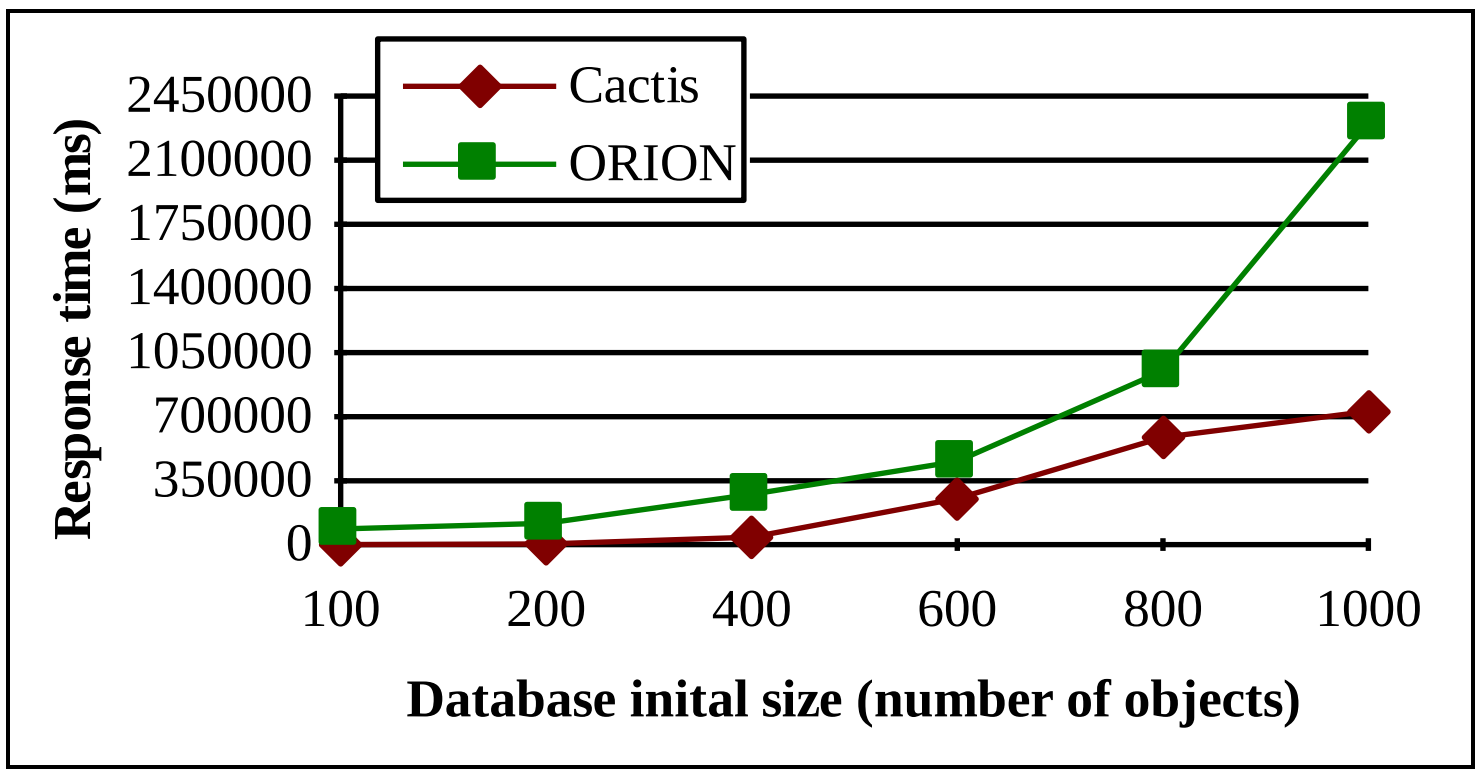

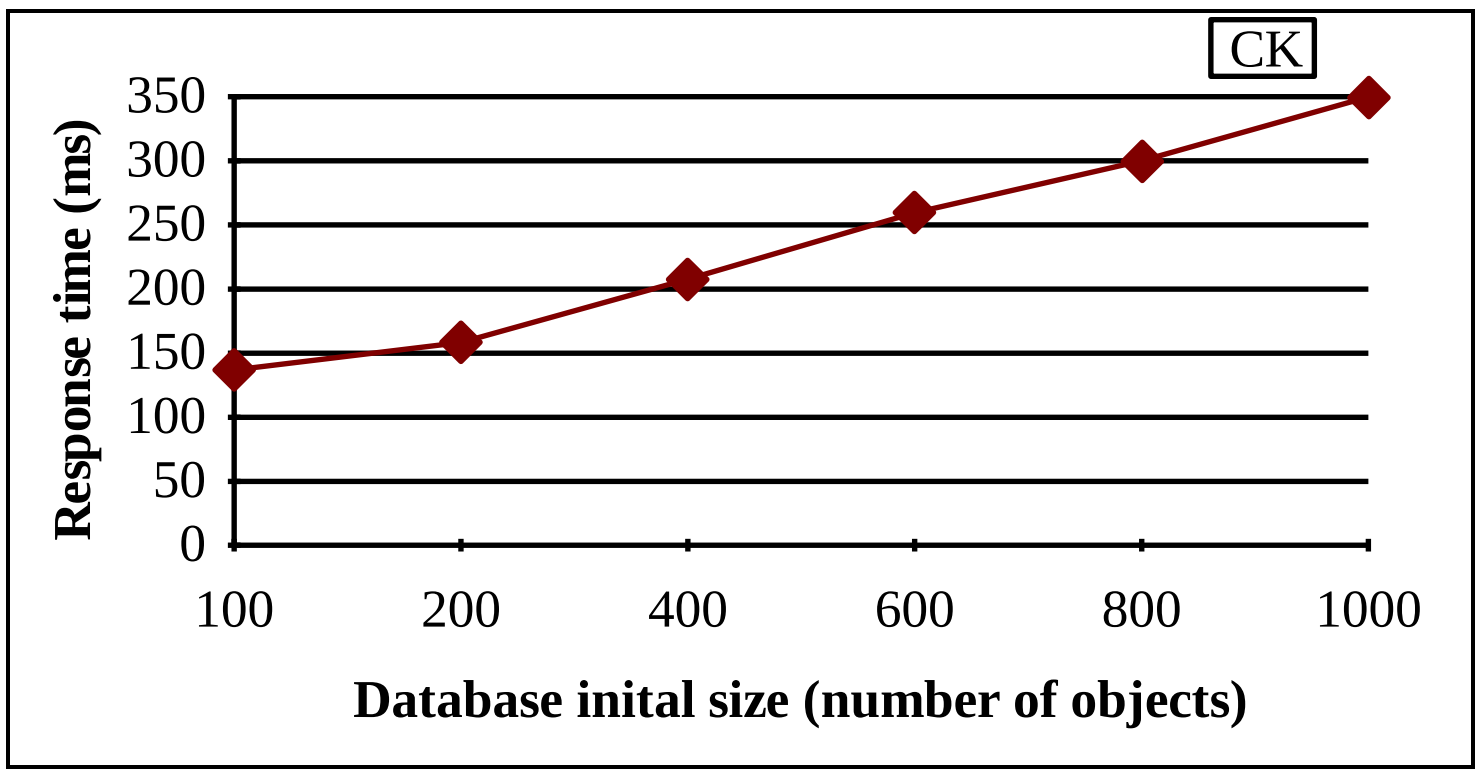

**Figure 10. Mean response time function of database initial size**

These results can be explained by looking at the mean number of I/Os (both transactions I/Os and clustering I/O overhead) function of the database initial size (Figures 11 and 12). Transactions I/Os giving an idea of how well a clustering algorithm places the objects, we can deduce from Figure 11 that objects are better clustered by CK and Cactis than by ORION (2.2 times better for Cactis). Cactis even appears to be slightly better (1.3 times) than CK.

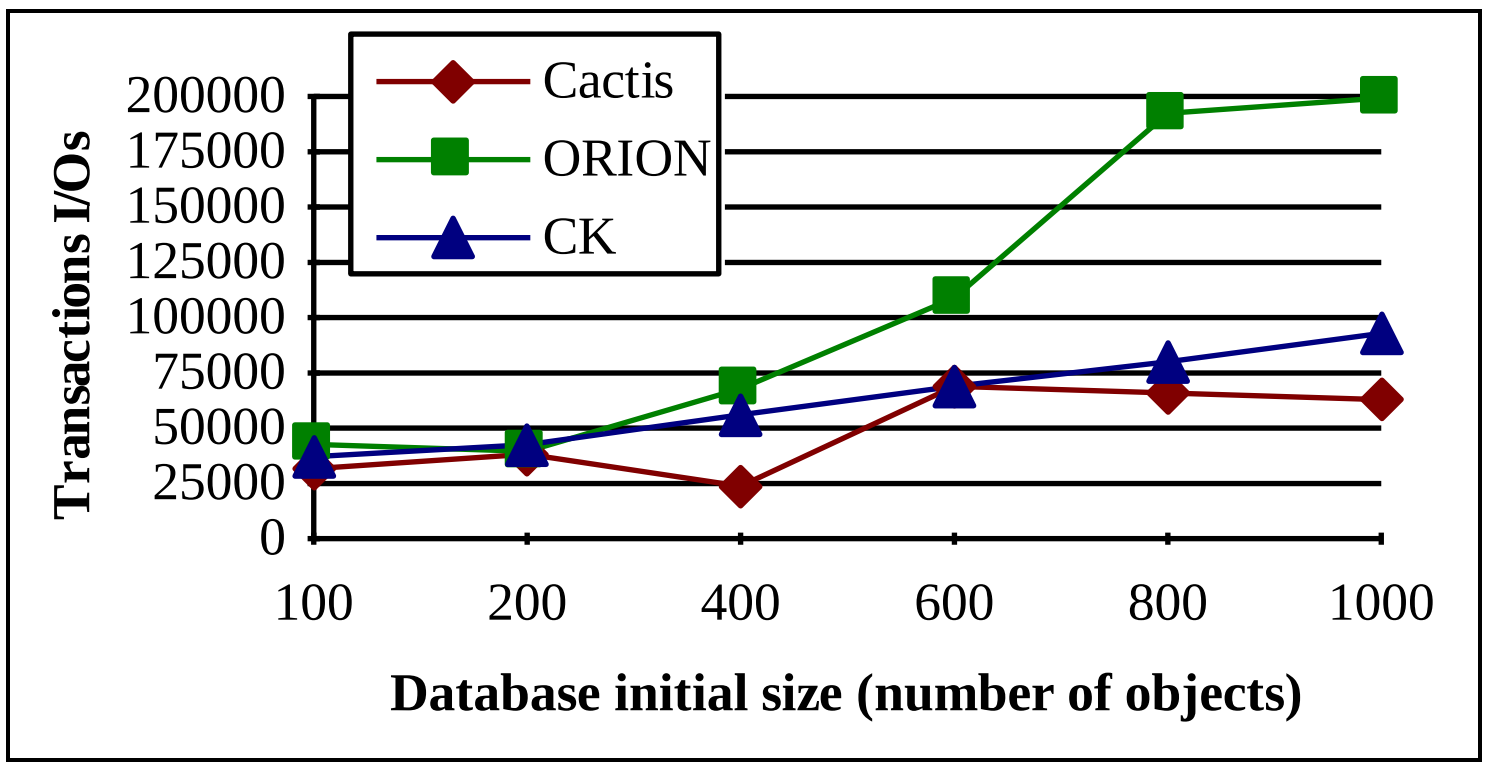

**Figure 11. Mean number of transaction I/O function of database initial size**

The fact that Cactis seems to cluster objects better than CK but shows worse overall performances can be explained by looking at Figure 12. They show that clustering overhead is 7,000 times greater for Cactis than for CK (clustering overhead for ORION being 1.4 times greater than for Cactis).

Such an outstanding performance is due to the true dynamic nature of CK, which is called only at object creation time and only accesses the object to cluster related objects, and not to the whole database as Cactis and ORION. Variations in clustering overhead come from variations in the number of created objects.

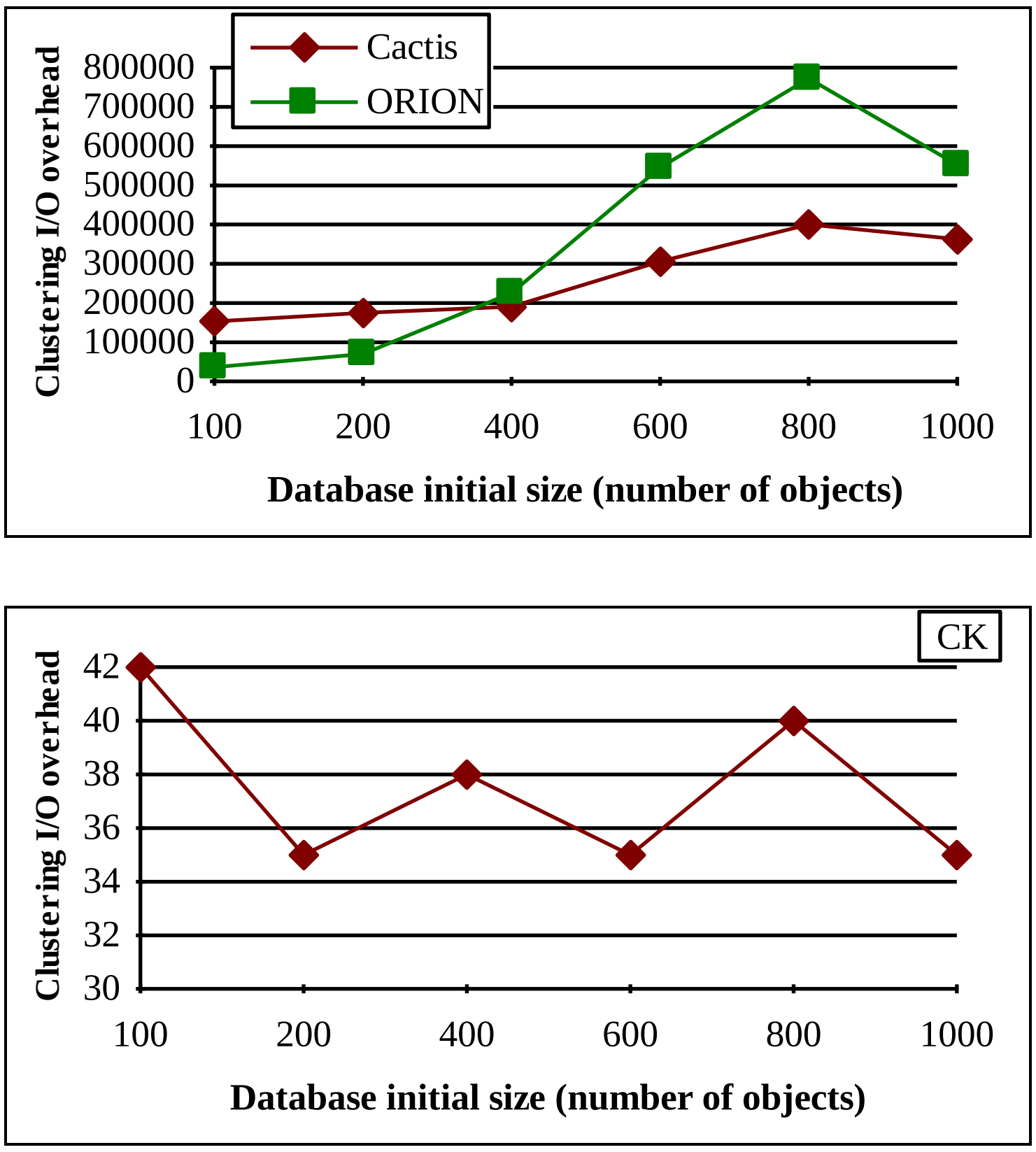

**Figure 12. Mean number of clustering I/O function of database initial size**

In terms of disk space, we expected the more sophisticated to use more space. Actually, the more a clustering algorithm is complex (i.e., the more it clusters object according to precise rules), the less a large number of objects are likely to share the same physical space (either page or segment). The mean number of disk pages used (Figure 13), as expected, is higher for the more complex algorithms, i.e., CK needs 1.7 times as many pages as Cactis and Cactis needs 2.8 times as many pages as ORION, for which the number of pages increases linearly.

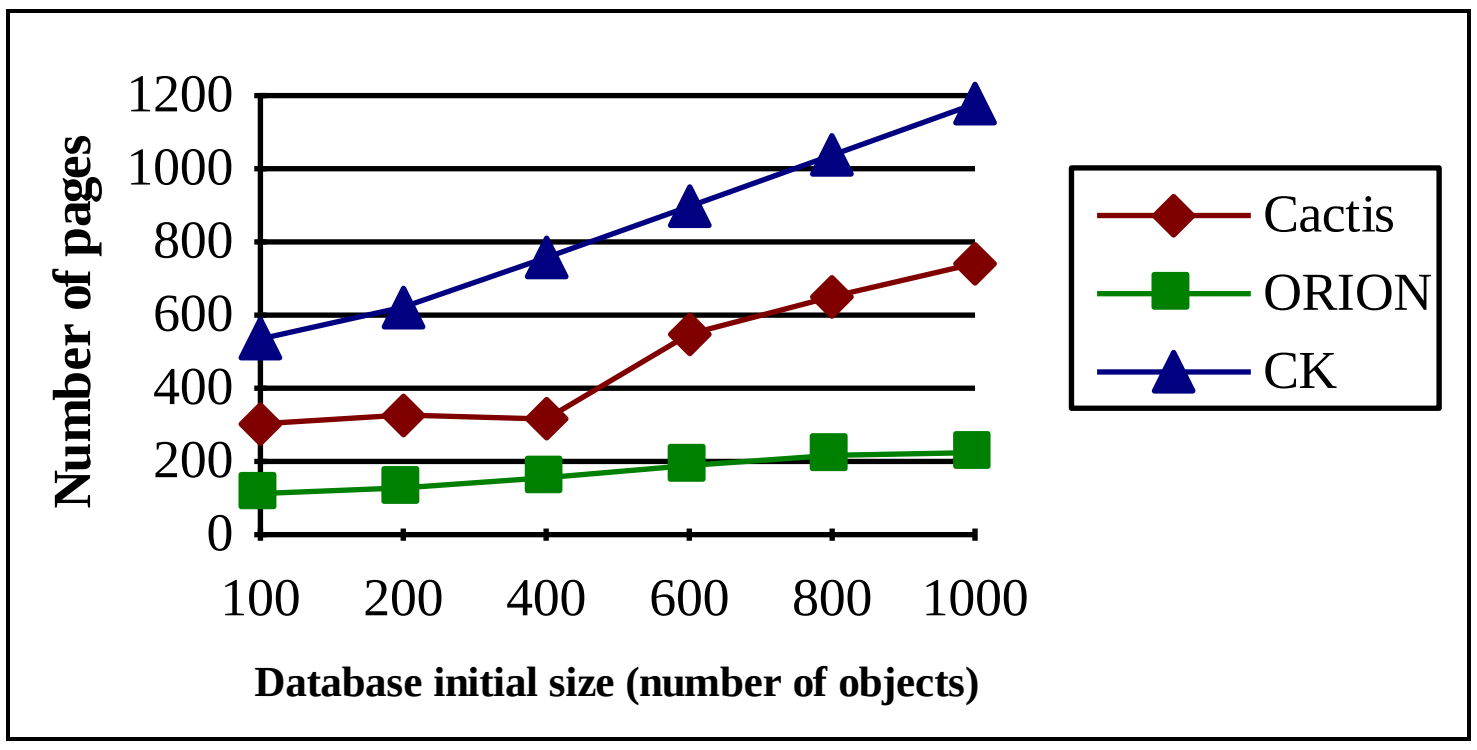

**Figure 13. Mean number of pages function of the database initial size**

### *5.2.2 Effects of the buffer capacity*

This series of simulations has been performed on a database of initial size 400 objects. By increasing the buffer capacity, we expect a decrease of the number of I/Os. As expected, Figure 14 shows that transactions I/Os decrease whatever clustering algorithm is used.

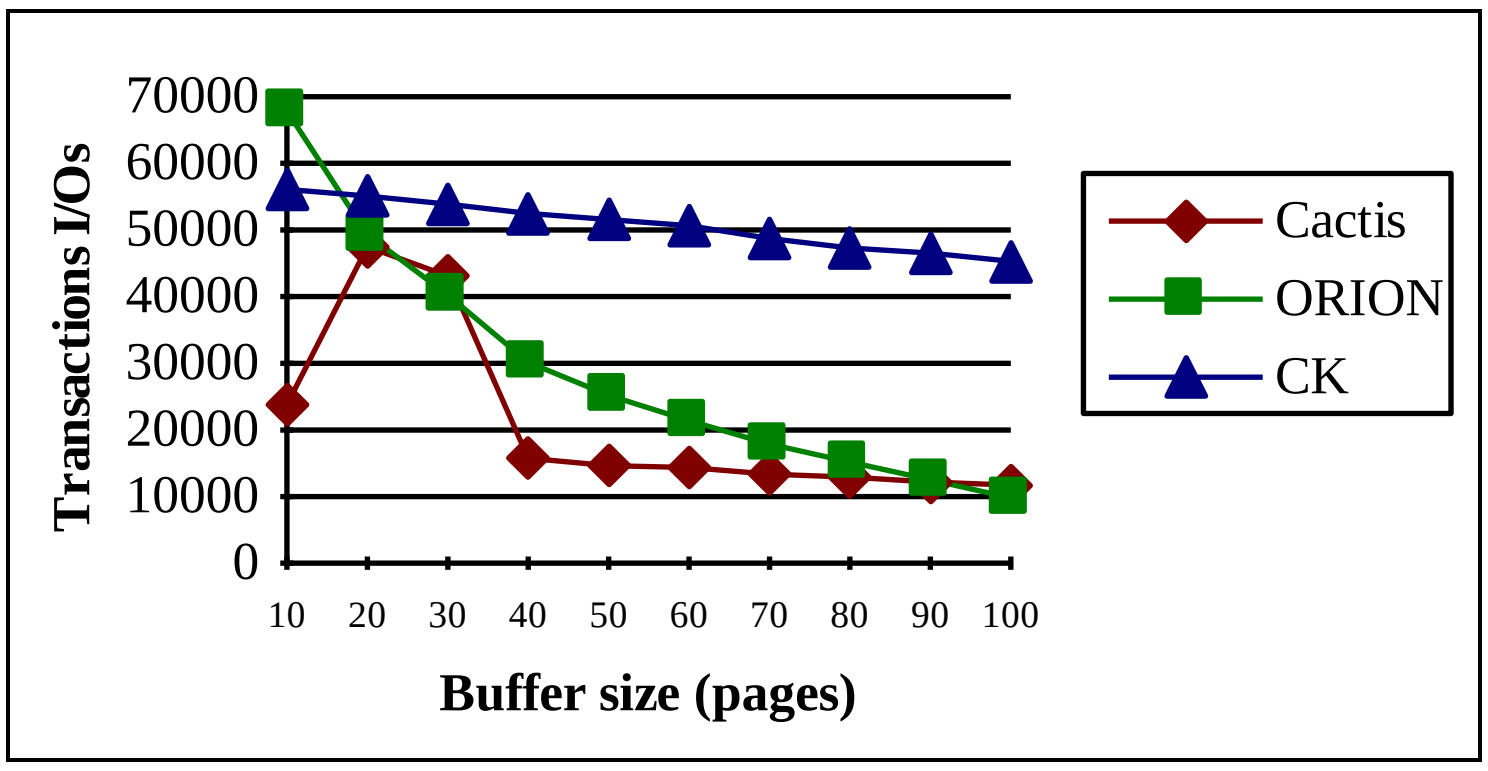

**Figure 14. Mean number of transaction I/O function of buffer capacity**

In the case of CK, the decrease is linear. In the case of Cactis, the number of I/Os decreases faster when the buffer capacity raises from 10 to 40 pages. Then it also becomes linear. The effect achieved with ORION is more spectacular. These results are due to the fact that ORION uses a smaller amount of pages than Cactis and Cactis uses a smaller amount of pages than CK to store the database. Thus, relatively to the database size, the buffer size increases faster for ORION than for Cactis and CK, hence allowing a greater and "faster" performance improvement. For instance, a buffer size of 20 pages represents 12 % of the database size for ORION against 6 % of the database size for Cactis and only 3 % of the database size for CK.

A decrease of clustering I/O overhead is also felt with a similar intensity (Figure 15) for Cactis and ORION because these algorithms scan all the database to reorganize it and thus take a great benefit from the increase in buffer capacity. In the case of CK, clustering overhead does not vary because too few objects are accessed each time for the increase in buffer capacity to be useful.

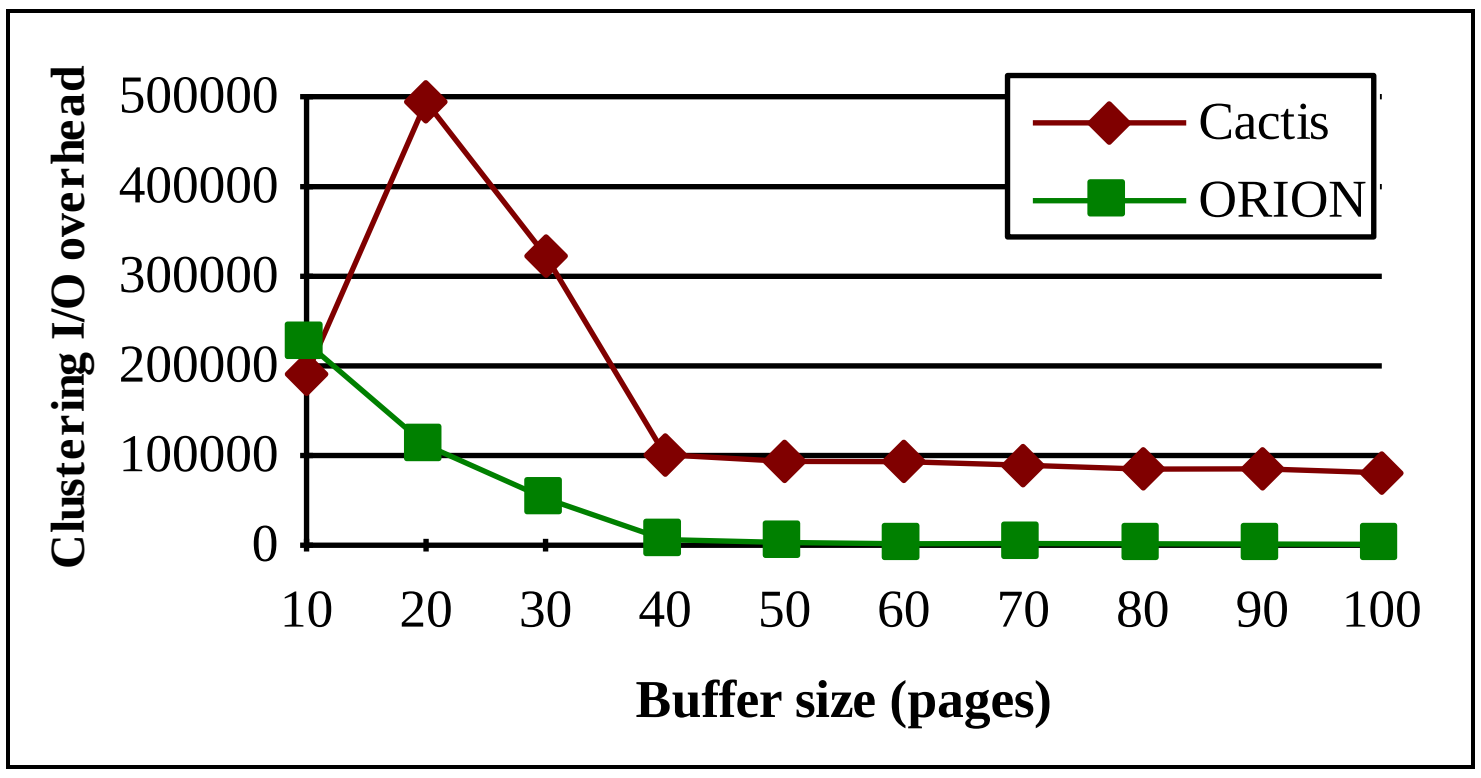

**Figure 15. Mean number of clustering I/O function of buffer capacity**

Figure 16 allows to measure in terms of global performance the relative performance improvements as the buffer capacity increases.

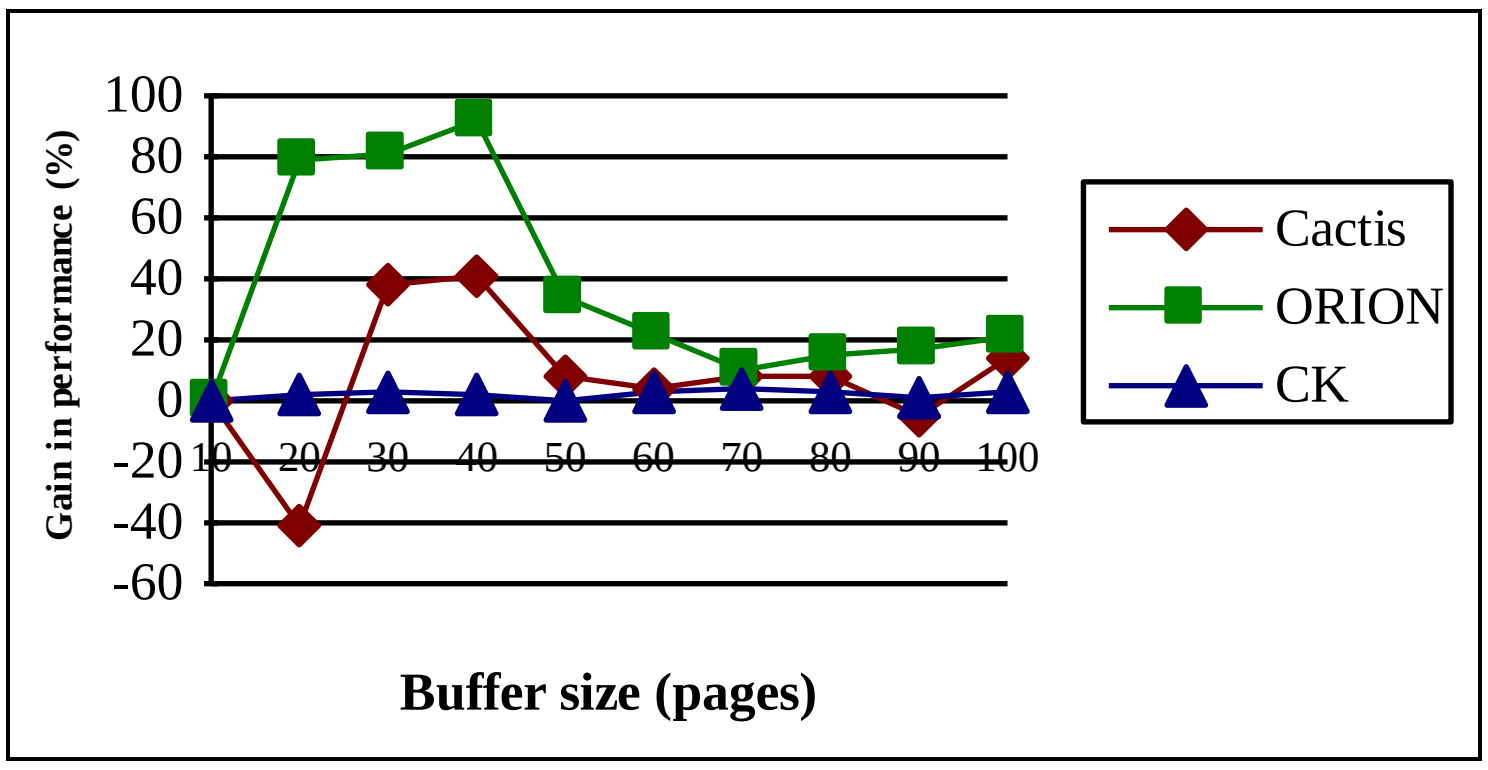

**Figure 16. Performance improvements function of buffer capacity**

We conclude that 50 pages seems to be the critical buffer size for Cactis and ORION (thus 14 % and 31 % of the database size, respectively). Beyond this critical size, performance improvements due to the increase in buffer capacity are lesser. CK performances are not significantly affected by variations in buffer size.

### *5.2.3 Effects of the read/write ratio*

Read/Write ratio is an important factor when seeking to evaluate DBMSs performances. Furthermore, [CHAN89] claims that CK algorithm performs better when the read/write ratio is high. For our simulation experiments, we used an initial database of 400 objects and a buffer size of 10 pages.

The performance of the Cactis and ORION algorithms decreases when the read/write ratio decreases. On the contrary, response time decreases along with the read/write ratio in the case of CK (Figure 17).

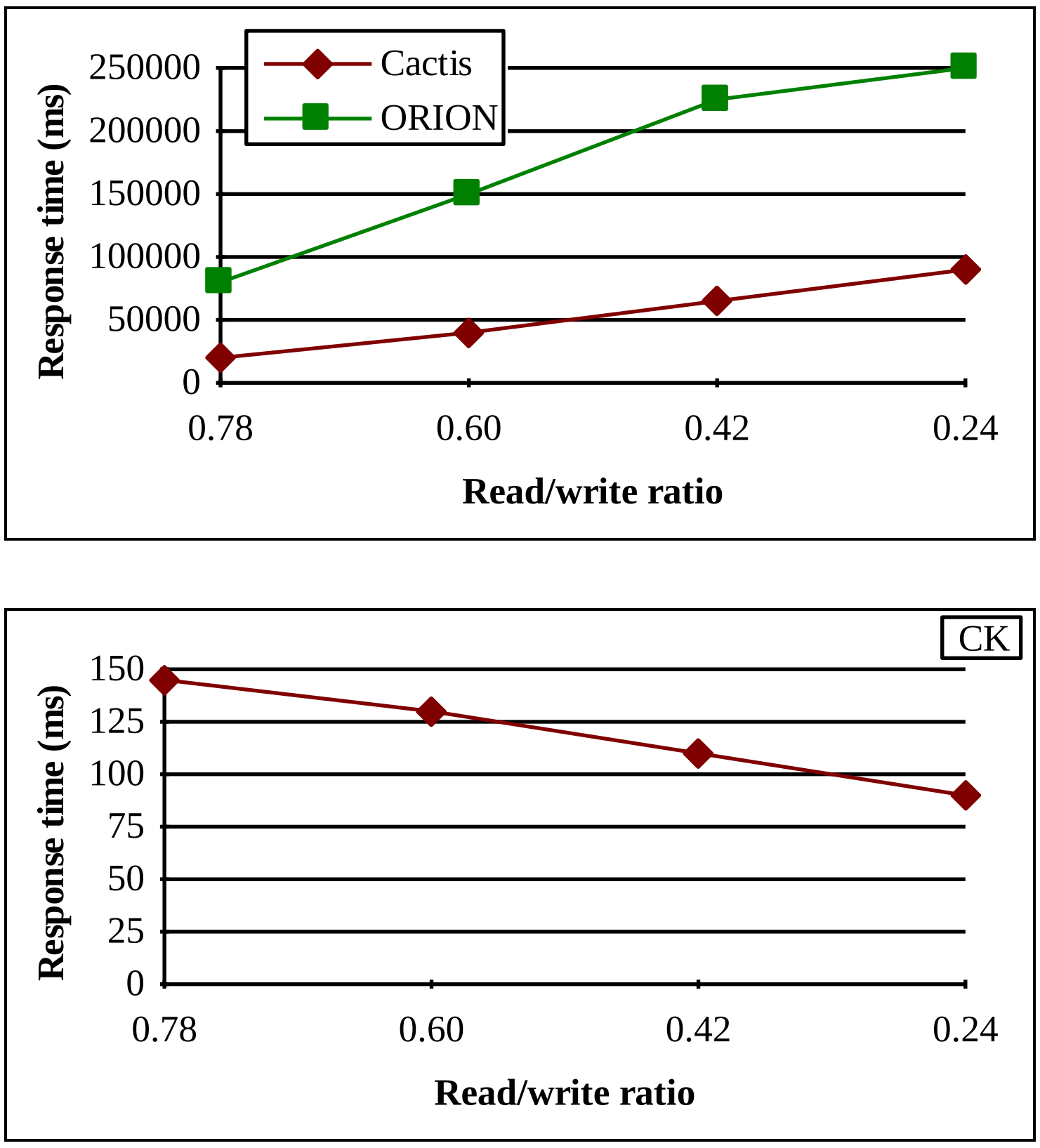

**Figure 17. Mean response time function of read/write ratio**

Since Object Creation is a write operation, the more the read/write ratio drops, the more the database size increases, thus implying more clustering overhead and confirming what is said in [CHAN89]. At the same time, transactions I/Os are slowly decreasing in number for Cactis and CK. This is because one single instance creation is less costly than, for instance, such queries as Q2: Range lookup or Q8: Sequential Scan. That explains the raise in performance for CK, since transactions I/Os drops from 10,000 to 5000 while clustering I/O overhead only rises from 100 to 500. In the Cactis case, clustering overhead is too important to compensate the decrease in transactions I/Os. For ORION, transactions I/Os increase anyway because of the poor clustering ability of the algorithm.

### *5.2.4 Impact of the query type on performances*

The queries whose types are presented in Section 3 access to objects according to different schemes. Hence, a clustering policy that is adapted to a certain type of query may not be adapted to another type. To evaluate the impact of the type of query on global performances, we measured transactions I/Os, only allowing each time one type of query to run. Results are summarized in Figure 18.

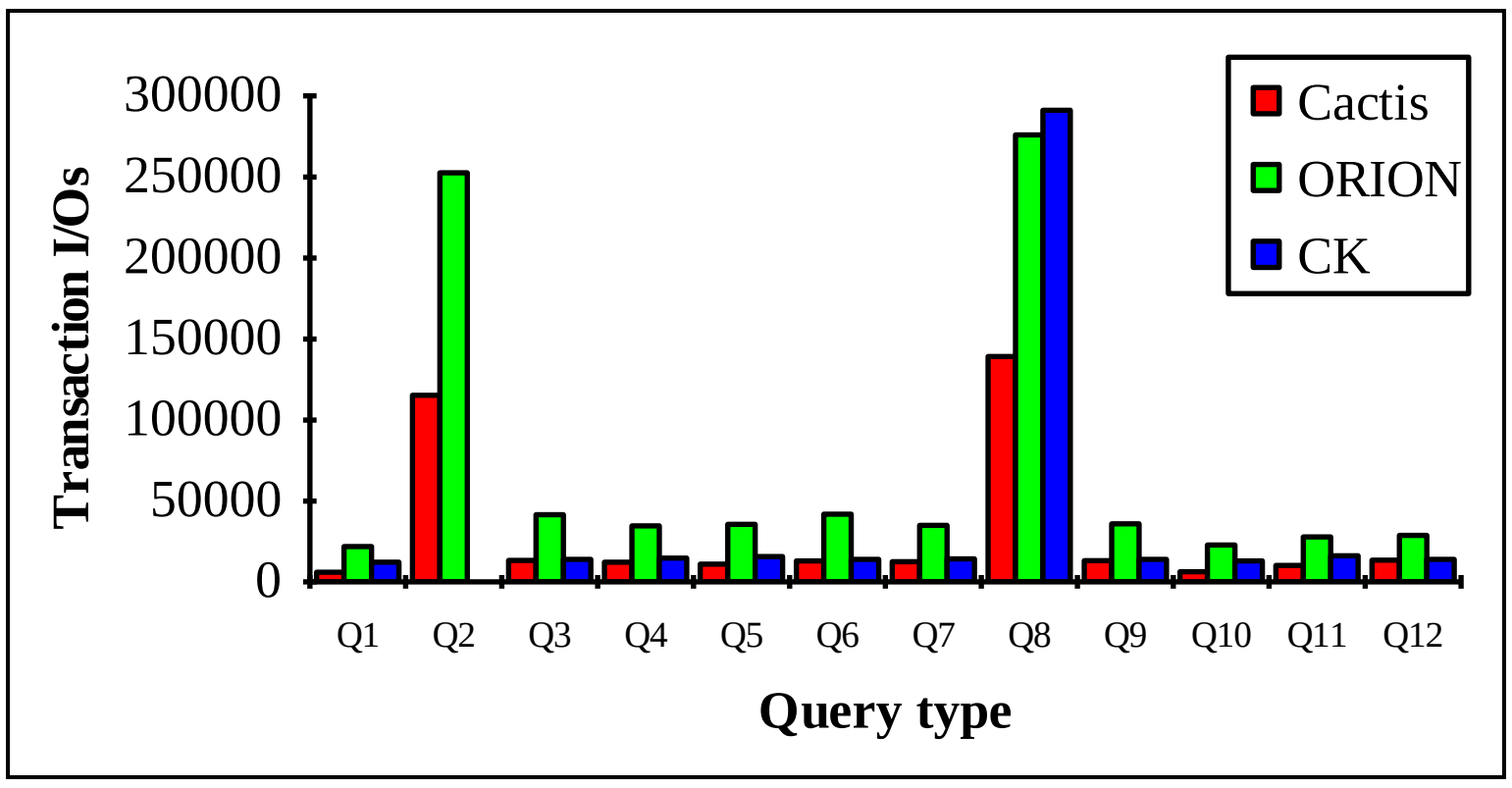

**Figure 18. Performances function of query type**

These results first confirm those obtained when evaluating clustering capabilities. It is also blatant that queries that access all the instances of a class (Q2: Range lookup and Q8: Sequential scan) do not benefit from clustering at all, whatever the clustering algorithm used.

## 6. CONCLUSIONS

Our simulation experiments clearly show that the CK algorithm outperforms both Cactis and ORION in terms of overall performance. This is due to both a good clustering capability and to the dynamic conception of the algorithm that allows an extremely low clustering overhead. Since the CK algorithm is activated only at object creation time and only accesses the few objects that are related to the newly created object, transactions are never blocked very long during clustering, as they are when the Cactis or the ORION algorithm is used. (The Cactis and ORION algorithms have to access all the objects in the database, even several times in the case of ORION, to reorganize the database; and transactions cannot be run when a reorganization occurs.) CK good clustering capability is based on the users' hints that specify the inter-objects access frequencies for each structural relationship and thus allows to cluster together objects that are likely to be accessed together.

Our simulations also showed that Cactis had a good clustering capability too, due to the use of statistics. Indeed, objects access frequencies and relationships use frequencies allow to cluster together objects that are actually accessed together. Though, the performances of the Cactis algorithm are greatly handicapped by clustering overhead that increases very quickly with the number of objects. However, this algorithm has been designed to run when the database is idle so that reclustering does not alter the database performance. Hence, if clustering overhead was not taken into account, the Cactis algorithm should perform about as well as CK algorithm as long as the statistics used during the last reorganization are pertinent.

In terms of disk space, the ORION algorithm is the less greedy algorithm. Then the Cactis algorithm follows, using almost half the number of disk pages needed by CK to cluster the database. However, when reorganizing the database, the Cactis and ORION algorithms need to build a new set of pages before deleting the old one. Thus they require about twice as much space as our graphs show. Hence, Cactis and CK are almost equivalent, ORION staying the less greedy algorithm in terms of disk space.

We have presented in this paper a methodology allowing the design of a tool enabling the *a priori* study or *a posteriori* comparison of the performances of clustering algorithms. This tool may be re-used since it is very easy to instance our generic action model with other clustering policies than those we chose to study. This tool may also be modified. It is particularly interesting in future developments to take into account buffering management strategies because it is mostly the use of both clustering and buffering techniques rather than clustering techniques alone that are found in the literature when speaking of performance improvement.

Our modelling methodology itself may also be re-used to model either another environment, or to build models designed to test the performances of other components of an OODB, or even to *a priori* model the global behavior of a DBMS in order to determine some management strategies to use.

## APPENDIX: CK CLUSTERING ALGORITHM [CHAN90]

```
PROCEDURE cluster_object(target_objet)
BEGIN
    /* step 1: get initial information */
    cluster_policy:=get_policy( );              /* Is page splitting enabled? */
    copy_set:=get_by_copy_set( );               /* Inherited attributes implemented by copy. */
    ref_set:=get_by_ref_set( );                 /* Inherited attributes implemented by reference. */
    inh_page_set:=get_all_inh_page( );          /* Source pages for inherited attributes. */
    struct_page_set:=get_all_struct_page( );    /* Source pages for structural objects. */
    page_set:=inh_page_set+struct_page_set;
    /* step 2: calculate ref_set lookup cost for each page */
    FOR p IN page_set                           /* If by-reference attribute r is */
        FOR r IN ref_set                        /* not in page p, storing target object */
            IF r NOT_IN p                       /* in page p requires one run-time */
            BEGIN                               /* lookup for attribute r. */
                weight(p):=1/(prob(p,struct_rel));
                Ref_LookUp(p):=Ref_LookUp(p)+weight(p);
            END;
    /* step 3: calculate copy_set lookup and storage cost for each page */
    FOR c IN copy_set                           /* If by-copy attribute c is not in page */
        FOR p IN page_set                       /* p, we could either cache it in page p */
            IF c NOT_IN p                       /* or change its implementation to be */
            BEGIN                               /* by-reference. */
                weight(p):=1/(prob(p,struct_rel));
                Copy_storage(p):=Copy_storage(p)+size_of(c);
                Copy_LookUp(p):=Copy_LookUp(p)+weight(p);
            END;
    /* step 4:  calculate total cost of every page. If by-copy attributes are */
    /*          implemented by reference, the total cost of storing target object */
    /*          in page p is represented by Total(p,1). Otherwise, the cost */
    /*          is represented by Total(p,2). */
    FOR p IN page_set
        Total_cost(p,1):=Ref_LookUp(p)*Lookup_cost+Copy_LookUp(p)*Lookup_cost;
        Total_cost(p,2):=Ref_LookUp(p)*Lookup_cost+Copy_storage(p)*Storage_cost;
    /* step 5: pick up best candidate page and try to insert the object */
    candidate_page:=Minimum(Total_cost);
    IF (cluster_policy EQ no_split)
        WHILE (NOT_FIT(candidate_page))
            candidate_page:=Next_Min(Total_cost);
    IF ((cluster_policy EQ page_split) AND (NOT_FIT(candidate_page))
        Split_page(candidate_page);
END;
```

## REFERENCES

[ANDE90], T.L. Anderson, A.J. Berre, M. Mallison, H.H. Porter III, B Scheider, "The HyperModel Benchmark", International Conference on Extending Database Technology, Venice, Italy, March 1990

[BANC92], F. Bancilhon, C. Delobel, P. Kanellakis, "Building an Object-Oriented Database System: The Story of $O_2$", Morgan Kaufmann Publishers, 1992

[BANE87], J. Banerjee, H.-T. Chou, J.F. Garza, W. Kim, D. Woelk, N. Ballou, H.-J. Kim, "Data Model Issues for Object-Oriented Applications", ACM Transactions on Office Information Systems, Vol. 5, No. 1, January 1987

[BERR91], A.J. Berre, T.L. Anderson, "The HyperModel Benchmark for Evaluating Object-Oriented Databases", in "Object-Oriented Databases with Applications to CASE, Networks and VLSI CAD", Edited by R. Gupta and E. Horowitz, Prentice Hall Series in Data and Knowledge Base Systems, 1991

[BULL95], F. Bullat, "Regroupement physique d'objets dans les bases de données", to appear in ISI, Vol. 3, No. 4, September 1995

[CATT91], R.G.G. Cattell, "An Engineering Database Benchmark", in "The Benchmark Handbook for Database Transaction Processing Systems", Edited by Jim Gray, Morgan Kaufmann Publishers, 1991

[CHAB93], S. Chabridon, J.-C. Liao, Y. Ma, L. Gruenwald, "Clustering Techniques for Object-Oriented Database Systems", 38th IEEE Computer Society International Conference, San Francisco, February 1993

[CHAN89], E.E. Chang, R.H. Katz, "Exploiting Inheritance and Structure Semantics for Effective Clustering and Buffering in an Object-Oriented DBMS", ACM SIGMOD International Conference on Management of Data, Portland, Oregon, June 1989

[CHAN90], E.E. Chang, R.H. Katz, "Inheritance in computer-aided design databases: semantics and implementation issues", CAD, Vol. 22, No. 8, October 1990

[CHEN76], D. Chen, "The Entity Relationship Model – Toward a Unified View of Data", ACM Transactions on Database Systems, March 76

[CHEN91], J.R. Cheng, A.R. Hurson, "Effective clustering of complex objects in object-oriented databases", ACM SIGMOD International Conference on Management of Data, Denver, Colorado, May 1991

[DARM95], J. Darmont, A. Attoui, M. Gourgand, "Performance Evaluation for Clustering Algorithms in Object-Oriented Database Systems", Springer Verlag Lecture Notes in Computer Science, DEXA 95 proceedings, London, September 1995

[FORD88], S. Ford, J. Joseph, D.E. Langworthy, D.F. Lively, G. Pathak, E.R. Perez, R.W. Peterson, D.M. Sparacin, S.M. Thatte, D.L. Wells, S. Agarwala, "ZEITGEIST: Database

Support for Object-Oriented Programming", 2nd International Workshop on Object-Oriented Database Systems, Bad Münster am Stein-Ebernburg, FRG, September 1988 [GOUR91], M. Gourgand, P. Kellert, "Conception d'un Environnement de Modélisation des Systèmes de Production", 3rd Industrial Engineering International Congress, Tours, France, March 1991

[GRUE91], L. Gruenwald, M.H. Eich, "MMDB Reload Algorithms", ACM SIGMOD International Conference on Management of Data, Denver, Colorado, May 1991

[HE93], M. He, A.R. Hurson, L.L. Miller, D. Sheth, "An Efficient Storage Protocol for Distributed Object-Oriented Databases", IEEE Parallel & Distributed Processing, 1993

[HUDS89], S.E. Hudson, R. King, "Cactis: A Self-Adaptive Concurrent Implementation of an Object-Oriented Database Management System", ACM Transactions on Database Systems, Vol. 14, No. 3, September 1989

[KIM90], W. Kim, J.F. Garza, N. Ballou, D. Woelk, "Architecture of the ORION Next-Generation Database System", IEEE Transactions on Knowledge and Data Engineering, Vol. 2, No. 1, March 1990

[SRIN91], V. Srinivasan, M.J. Carey, "Performance of B-Tree Concurrency Control Algorithms", ACM SIGMOD International Conference on Management of Data, Denver, Colorado, May 1991

[TSAN91], M.M. Tsangaris, J.F. Naughton, "A Stochastic Approach for Clustering in Object Bases", ACM SIGMOD International Conference on Management of Data, Denver, Colorado, May 1991

[TSAN92], M.M. Tsangaris, J.F. Naughton, "On the Performance of Object Clustering Techniques", ACM SIGMOD International Conference on Management of Data, San Diego, California, June 1992